\documentclass{aa}
\usepackage{psfig}
\usepackage{natbib}
\usepackage{graphicx}

\begin{document}

\def\eps{\varepsilon}
\def\aap{A\&A}
\def\apj{ApJ}
\def\apjs{ApJ Supp.}
\def\apjl{ApJL}
\def\mnras{MNRAS}
\def\aj{AJ}
\def\nat{Nature}
\def\aaps{A\&A Supp.}
\def\prd{Phys. Rev. D}
\def\prl{Phys. Rev. Lett.}
\def\PR{Phys. Rept.}
\def\NIMA#1#2#3{#2, {\rm Nucl.~Instr.~Methods {\bf A}}, #1, #3}
\def\PLB#1#2#3{#2,  {\rm Phys.~Lett. {\bf B}}, #1, #3}
\def\PRB#1#2#3{#2,  {\rm Phys.~Rev. {\bf B}}, #1, #3}
\def\PRD#1#2#3{#2,  {\rm Phys.~Rev. {\bf D}}, #1, #3}

\def\PR#1#2#3{#2,  {\rm Phys.~Rept.}, #1, #3}
\newcommand{\vt}{\mbox{\bf {T}}}
\newcommand{\vtcmb}{\mbox{\bf {T}}_{cmb}}
\newcommand{\vm}{\mbox{\bf {M}}}
\newcommand{\vn}{\mbox{\bf {N}}}
\newcommand{\vf}{\mbox{\bf {F}}}
\newcommand{\1}{\'\i}
\newcommand{\Au} {\mbox{$ ^{197}{\rm{Au}}$}~}
\newcommand{\Aus} {\mbox{$ ^{197}{\rm{Au}}$}}
\newcommand{\neut}{$\tilde{\chi}$~}
\newcommand{\neu}{$\tilde{\chi}$}
\newcommand{\Co} {\mbox{$ ^{57}{\rm{Co}}$}~}
\newcommand{\Cos} {\mbox{$ ^{57}{\rm{Co}}$}}
\newcommand{\GeVc}    {\mbox{$ {\mathrm{GeV}}/c                            $}}
\newcommand{\GeVcc}    {\mbox{$ {\mathrm{GeV}}/c^2                           $}}
\newcommand{\MeVc}    {\mbox{$ {\mathrm{MeV}}/c                            $}}
\newcommand{\hetrois}    {\mbox{$ ^{3}{\mathrm{He}}                            $}~}
\newcommand{\hetro}    {\mbox{$ ^{3}{\mathrm{He}}                            $}}
\newcommand{\tritium}    {\mbox{$ ^{3}{\mathrm{H}}                            $}~}
\newcommand{\hequatre}    {\mbox{$ ^{4}{\mathrm{He}}                            $}~}
\newcommand{\mydeg}   {\mbox{$ ^\circ                                      $}}
\newcommand{\DM}{Dark Matter}
\newcommand{\gam}{\mbox{\rm $\gamma$-ray}~} 
\newcommand{\gams}{\mbox{\rm $\gamma$-rays}~}
\newcommand{\gamss}{\mbox{\rm $\gamma$-rays}}
\newcommand{\ou}{o\`u}
\newcommand{\elec}{$e^{-}$}
\newcommand{\posi}{$e^{+}$}
\newcommand{\aprot}{\={p}}
\newcommand{\micron}{\mbox{{\rm$\mu$m}}}
\newcommand{\microns}{\mbox{{\rm$\mu$m}}~}
\newcommand{\muK}{\mbox{{\rm $\mu$K}}~}
\newcommand{\muKs}{\mbox{{\rm $\mu$K}}}

\def\me{m_\e}
\def\lesssim{\mathrel{\hbox{\rlap{\hbox{\lower4pt\hbox{$\sim$}}}\hbox{$<$}}}}
\def\gtrsim{\mathrel{\hbox{\rlap{\hbox{\lower4pt\hbox{$\sim$}}}\hbox{$>$}}}}

\def\vr{\vec{r}}
\def\vrp{\vec{r}_\perp}

\def\del#1{{}}

\def\C#1{#1}

\input epsf
\def\plotancho#1{\includegraphics[width=17cm]{#1}}

\title{An analysis method for time ordered data  processing of  
Dark Matter experiments}
\titlerunning{An analysis method for Dark Matter data processing}

\author{E. Moulin\inst{1}\thanks{Present address : LPTA Montpellier, place E. Bataillon, 34095 Montpellier cedex 5 (France) }
    \and J.~F. Mac\'{\i}as-P\'erez\inst{1}   
    \and F. Mayet\inst{1}       \and \\
     C. Winkelmann\inst{2}
     \and  Yu.~M.~Bunkov\inst{2}
     \and H. Godfrin\inst{2}
      \and D. Santos\inst{1} }

\institute{
Laboratoire de Physique Subatomique et de Cosmologie, 
 CNRS/IN2P3 et Universit\'e Joseph Fourier, 
 53, avenue des Martyrs, 38026 Grenoble cedex, France 
\and
Centre de Recherches sur les Tr\`es Basses Temp\'eratures,
 CNRS et Universit\'e Joseph Fourier, BP166, 38042 Grenoble cedex 9, France
}

\date{\today}

\mail{Emmanuel.Moulin@lpta.in2p3.fr}

\abstract{The analysis of the time ordered data  of  
Dark Matter experiments is becoming more and more challenging with the increase
of sensitivity in the ongoing and forthcoming projects. 
Combined with the well-known level of background events, this leads to a rather high
level of pile-up in the data. Ionization, scintillation as well as bolometric signals present
common features in their acquisition timeline: low frequency baselines, random gaussian noise, 
parasitic noise and signal characterized by well-defined peaks.  
In particular, in the case of long-lasting signals such as bolometric ones,
the pile-up of events may lead to an inaccurate reconstruction of the
physical signal (misidentification as well as fake events).}
{We present a general method to detect and extract signals in noisy data with 
a high pile-up rate and qe show that events from few keV to hundreds of keV can be reconstructed 
in time ordered data presenting a high pile-up rate.}
{This method is based on an iterative detection and fitting procedure
combined with prior wavelet-based denoising of the data and baseline subtraction.}
{We have tested this method on simulated data of the MACHe3 prototype experiment
and shown that the iterative fitting procedure allows us to recover the
lowest energy events, of the order of a few keV, in the presence of background
signals from a few to hundreds of keV. Finally we applied this method to the
recent MACHe3 data to successfully measure the spectrum of conversion electrons
from \Co source and also the spectrum of the background cosmic muons.}
{} 

\keywords{Cosmology : Dark Matter - Methods : data analysis - numerical}
\maketitle

\section{Introduction \label{sec:intro}}
Evidence for the existence of non-baryonic Dark Matter seems to be well established from the
results of recent cosmological observations : 
CMB anisotropy measurements \cite{wmap,archcosmo,rebolo,boomerangcosmo}, 
large scale structure surveys \cite{lss} and type Ia supernova measurements \cite{sn1a}. The leading candidate 
intended to composed non-baryonic dark matter,
is the lightest 
neutralino \neut proposed by supersymmetric extensions of the standard model of Particle Physics \cite{jungman}.
A great amount of non-baryonic Dark Matter 
detectors \cite{cdms,cresst,edel} have been developed using different and complementary techniques 
to access to the energy release.
The related signals which are measured, can be ionization, scintillation or heat released by the particle interaction.
The simultaneous measurement of a combination of at least two of the latter quantities are used in order to achieve a 
high rejection factor against background events and systematic effects.\\

\hetrois has been proposed as a sensitive medium \cite{mayetnim,santosdm20} to overcome well-known limitations 
of existing Dark Matter detectors, such as for example the
discrimination between  WIMPs and background neutrons. Moreover such a nucleus will be sensitive to the
axial interaction of the neutralino \neu,
making a \hetro-based detector complementary to most of existing detectors~\cite{susy1,susy2}.\\
Following early works \cite{pickett,bunkov,bauerle} on a \hetrois superfluid cell at 
ultra low temperature $\rm \sim 100 \mu K$, a 3-cell prototype, MACHe3 (MAtrix of Cells of \hetro), has
 been developed. It has been used to demonstrate for the first time the possibility 
 to measure events in the keV energy range in \hetrois \cite{electrons}.\\
 
As for many other bolometric detectors, e.g. Edelweiss \cite{edel}, Archeops \cite{tristram,pipeline}, the raw data of MACHe3
consist of time ordered samples with four main components :  1) A baseline coming mainly from low frequency fluctuations
of the thermal bath used to cool down the detector. 2) Random Gaussian noise. 3) Parasitic noise characterized by medium and high frequency structures. 
4) The signal of interest convoluted by the bolometer time response. The latter
when related to the energy deposited by incoming particles leads to peaks in the data. In this paper we present the method developed
for the analysis of these data \cite{these}. In particular we focus on the detection of peaks in noisy data presenting a high level
of pile-up. The method and algorithms used for this analysis are general and 
may well be used for other applications.

This paper is organized as follows. Section~\ref{sec:data} presents the main characteristic of 
typical bolometric data. In section~\ref{sec:anal} 
we deal with the details of the processing procedure. Sections \ref{sec:appli.simu} and 
\ref{sec:appli.mache3} presents the main results obtained after processing
of the simulated and of the MACHe3 data respectively. Finally, we conclude in section~\ref{sec:conc}.

\section{Description of the time ordered data \label{sec:data}}
We present in this section the main characteristics of the MACHe3 experiment
and of its raw time ordered data which
as described above are shared by other bolometric detectors.
 
\subsection{Experimental set-up}

A multi-cell prototype of MACHe3, with three bolometer cells (A, B and C) aligned longitudinally, has been developed 
and operated at the nominal temperature of $\rm 100\ \mu K$ \cite{electrons}. 
The time constant for thermal relaxation of the cell is of the order of $\rm 3 \ s$
Each cell contains one vibrating wire resonator 
which is used to measure
the energy released \cite{bauerle,trique}. \\

A \Co radioactive source, which provides low energy electrons, 
was spot-welded in the inner wall of the center cell (B). 
Because of the short free path of electrons in \hetro, $\rm \leq 370 \ \mu m $ for an energy below $\rm 40 \ keV$, the energy 
released by the interacting electrons will be fully detected within
that cell. Further because of the relatively high activity of the source, for Dark Matter concern, about 0.06 Bq,  
we expect conversion electrons to pile-up at the 
experiment sampling rate of 100 ms.
In addition to electrons,   events corresponding to cosmic muons 
and $\gamma$-rays of 14.4, 122 and 136 keV
emitted by the source  are expected
in the three cells.

\subsection{Example of raw data at 100 \muK}
\begin{figure}[t]
\begin{center}
\includegraphics[scale=0.55]{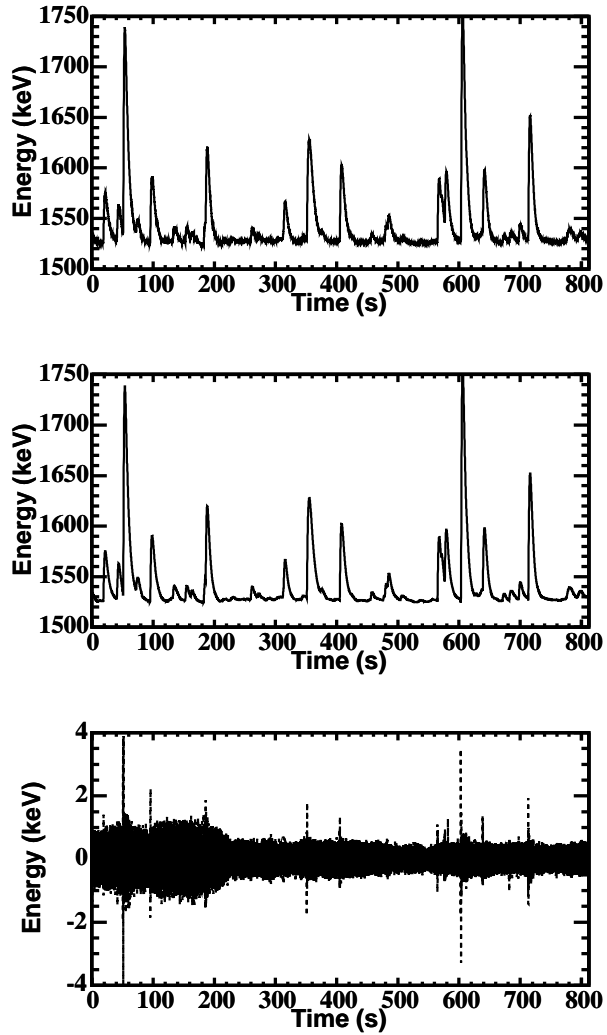}
\caption{Top: raw data from the central cell of the MACHe3 prototype, corresponding to the 
energy released   (in $\rm keV$)   
as a function of time.
Events are coming from cosmic muons and low energy conversion electrons from a
$\rm ^{57}Co$ source spot-welded on the inner wall of
the cell. Microvibrations as 
well as statistical noise add up to the baseline fluctuations. The sampling rate is 100 ms. 
Middle: same data but after applying the denoising algorithm.
 Bottom:  Residuals between raw and denoised data on an expanded scale. See text for details.}
\label{fig:rawdenoiseddata}
\end{center}
\end{figure}
\label{sec:raw}

The raw data consists of a sequence of peaks corresponding to the   energy released
by  interacting particles inside the cell. The peaks are characterized by
a rising and a relaxing time. The first one 
is related to the quality factor of the wire  (Q$\sim$10$^4$) whereas  the second depends on the geometry of the cell and 
the size of the orifice. 
The upper plot on figure~\ref{fig:rawdenoiseddata} shows a typical acquisition
spectrum with $\rm 100 \ ms$ sampling rate from the cell containing the \Co source (B cell). 
We can observe a baseline in data corresponding to low frequency fluctuations of the thermal bath. 
Cosmic muons as well as 
very low energy events corresponding to electrons coming from the source
are visible. Given the relatively high activity of \Co source we observe a high level of pile-up
in the data corresponding to multiple electron events, Auger as well as conversion, and also to combinations of
muons and electrons. The pile-up needs to be carefully treated in the analysis for
an accurate reconstruction of the low energy events. \\

Notice that peaks in the data
whatever their origin (muons, $\gamma$-rays, electrons) are 
identical in shape (same characteristic time constants). 
This is why, for the purpose of the 
analysis, we will define in the following a reference peak with only one free parameter: 
the amplitude which is given by the amount of energy released inside the cell.

\section{Data analysis method \label{sec:anal}}
We have developed an analysis method in four main steps. First of all,
we reduce significantly the statistical noise in the data using
a denoising algorithm. Second, the low frequency baseline
in the data is removed as estimated from the local minima. Then,
we estimate the shape of the bolometer time response by extracting 
a reference peak from the data. Finally, we apply to the data
an iterative fitting procedure taking as model the reference peak.
\subsection{Data denoising}
The time ordered data are processed in pieces
each consisting of 8192 samples. To reduce
the statistical noise, a wavelet-based denoising method is applied to data. 
This procedure is based on a shrinkage algorithm with conditional means
as described by~\cite{percival}. It presents two free parameters $j_{max}$ and
 $p$ :
\begin{itemize}
\item  $j_{max}$ is related to  the maximal wavelet scale at which
the denoising is applied. This implies that data are not modified at higher
scales. This parameter is chosen to keep both the shape and the amplitude of the
peaks and reduce significantly the noise. In other words, we focus in the
high frequency noise and we do not modify the low frequency components in the
data. 
\item $p$ expresses the
quality required for the denoising and is fixed to 0.9 for our analysis.
This value was chosen empirically but in general for values between 
0.8 and 0.95 the results present not significant differences.
\end{itemize}
In the middle plot of figure~\ref{fig:rawdenoiseddata} we present the data 
after applying the denoising algorithm.   We can observe that the method allows us  
to preserve both the shape and amplitude of the events. This is indicated by the
residuals between raw and denoised data plotted on the bottom on an expanded scale.
Residual amplitudes are about 1 keV or less. Shapes of the peaks are unchanged. 
Typically for a 200 keV peak, the residual is at most of
the order of 4 keV.

\subsection{Baseline subtraction}

We clearly observe in the bolometric time ordered data a baseline related to the
low frequency drifts of the thermal bath temperature plus a constant value.
Due to the high level of pile-up in the data, the determination and subtraction
of this baseline is difficult. An standard direct fit of this baseline is highly
biased by the presence of the signal peaks and the baseline is poorly
reconstructed.
In the same way, the application of another standard technique based on the localization and
subtraction of the peaks reduces too much the data to which the fitting procedure
can be applied.\\
After careful analysis of the properties of this baseline, we realized that it can be defined 
by the local minima in the data as the peaks from particle interactions
do not contribute to these minima. We assume that the noise distribution
is double-tailed and the peak signal strictly positive. 
In practice, we search for local minima in intervals of
812 samples making a total of 10 minima in the 8192 channel sample. 
The identified minima computed in this way are then fitted to a polynomial. We have 
found that in 8192 channel sample, a first order polynomial reproduces very well the baseline,
as temperature fluctuations of the \hetrois \ bath are at very low frequency. 
The polynomial fit is then subtracted from the data 
preserving both the shape and the amplitude of the peaks.   

\subsection{Reference peak}
\begin{figure}[ht]
\begin{center}
\includegraphics[scale=0.45]{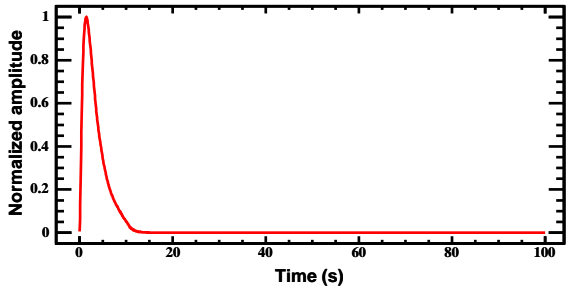}
\caption{Reference peak for data processing. This peaks was
extracted directly from the data (see text for details)
and completed   from time equal to 8 s,  by a decreasing exponential  to mimic at best
the time response of the bolometer.  The sampling rate is 100 ms.} 
\label{fig:referencepeak}
\end{center}
\end{figure}

As pointed out previously, a particle interaction produces a peak characterized by a rising and decreasing time (see section~\ref{sec:raw}) which depends only on the instrumental set up and 
does not depend on the amplitude of the signal as long as the
detector does not saturate. Because of this, it is useful 
to define a normalized reference peak (with an amplitude of 1) which will mimic the shape of the peaks observed
in the data and which can be used for further analysis of these data (see 
section~\ref{subsec:ite}).
To mimic at best the signal, we directly select the reference peak from the data with
no modeling. The chosen candidate must not be affected by pile-up and must
have a high signal to noise (amplitude versus r.m.s. noise) to 
limit the impact of the noise contribution. In practice, we
have chosen an isolated peak with signal to noise ratio larger than 10
which is normalized to an amplitude of unity. 
To ensure a high quality posterior analysis the peak is then completed   from time equal to 8 s, 
with a decreasing exponential function over a total of 1024 data samples. In this way we account
for the high end of the tail of the bolometric time response which is below the noise level for
the acquisition data.
Figure~\ref{fig:referencepeak} shows the reference peak extracted from the 
data and which was used for the data processing presented below. 

\subsection{Iterative fitting procedure \label{subsec:ite}}
We have developed an iterative fitting procedure to extract the
energy released in the bolometric cells by the interacting particles.
This procedure is based on three main steps.

\begin{enumerate} 
\item 
We search for the peaks on the data and flag them as detected. For this purpose, 
we compute the first derivative of the data. Then we select, on this new
data set, all samples for which the amplitude of the signal is five times
larger than the r.m.s noise (i.e. all  samples with a signal to noise
ratio better than five). 
  This allows us to get intervals in which a peak will be fitted. 
For a single peak we may flag various samples
which in some cases may be unconnected. To avoid fitting multiple peaks
when only a single one is detected but still to allow for
the detection of piled-up peaks, we broaden the flags over 35 samples. This is
about twice the rising time of the bolometer response.
\item 
We fit to the data as many model-peaks as peaks where flagged in the previous step,
  with a minimization of the mean square deviation between the data and the fit.
The free parameters in the fit are the position and the amplitude of the peak.
Each model-peak is constructed from the reference peak with initial position and
amplitude given by the position and value of the maximum of the signal within the 
corresponding flagged peak.   It is important to note that the peaks detected at the previous iteration
are refitted in the current iteration. It leads to a better fit for these peaks as contributions of 
other peaks detected during the actual iteration can be taken into account. 

\item 
We subtract the fitted model-peaks from the original data to estimate the residuals on the fitting procedure. 
If the fit was perfect, the residuals would directly correspond to the noise in the
data. Therefore, we can define, for each fitted peak, a signal to noise ratio ($\rm S/N$) by dividing
the amplitude of the fitted peak by the maximum of the absolute value of the residuals
over 35 samples centered at the position of the 
maximum\footnote{It has to be noted that the signal over noise ratio defined in this step
is different from the one calculated in the first step from the derivative of the data.}. Finally, each fitted peak is flagged
and we take these flags and the residuals as inputs for the next iteration step.
This way we can search for smaller amplitude peaks keeping a trace of the larger
amplitude ones.
\end{enumerate}
\begin{figure}[t]
\begin{center}
\mbox{\includegraphics[scale=0.48]{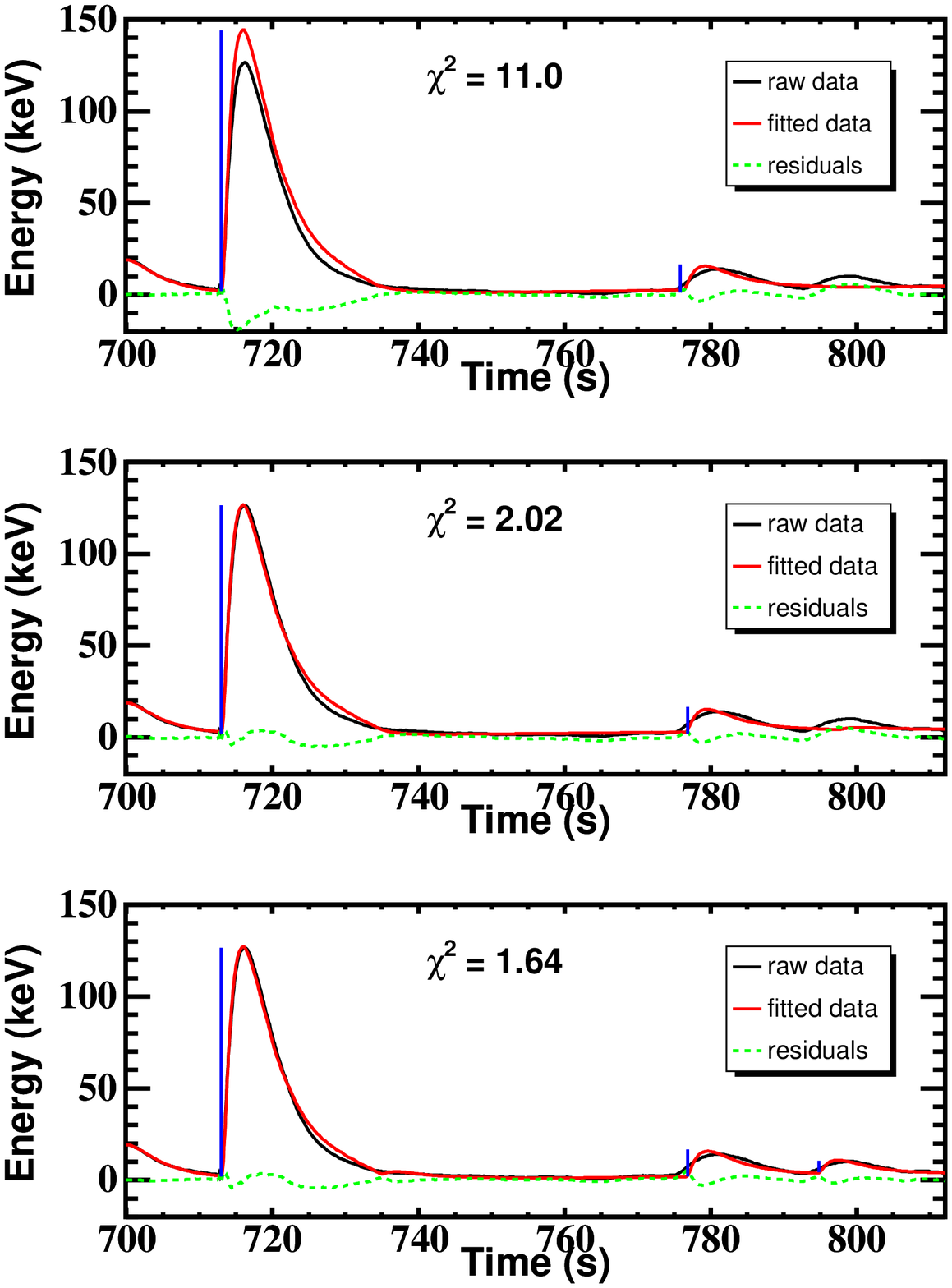}}
\mbox{\includegraphics[scale=0.48]{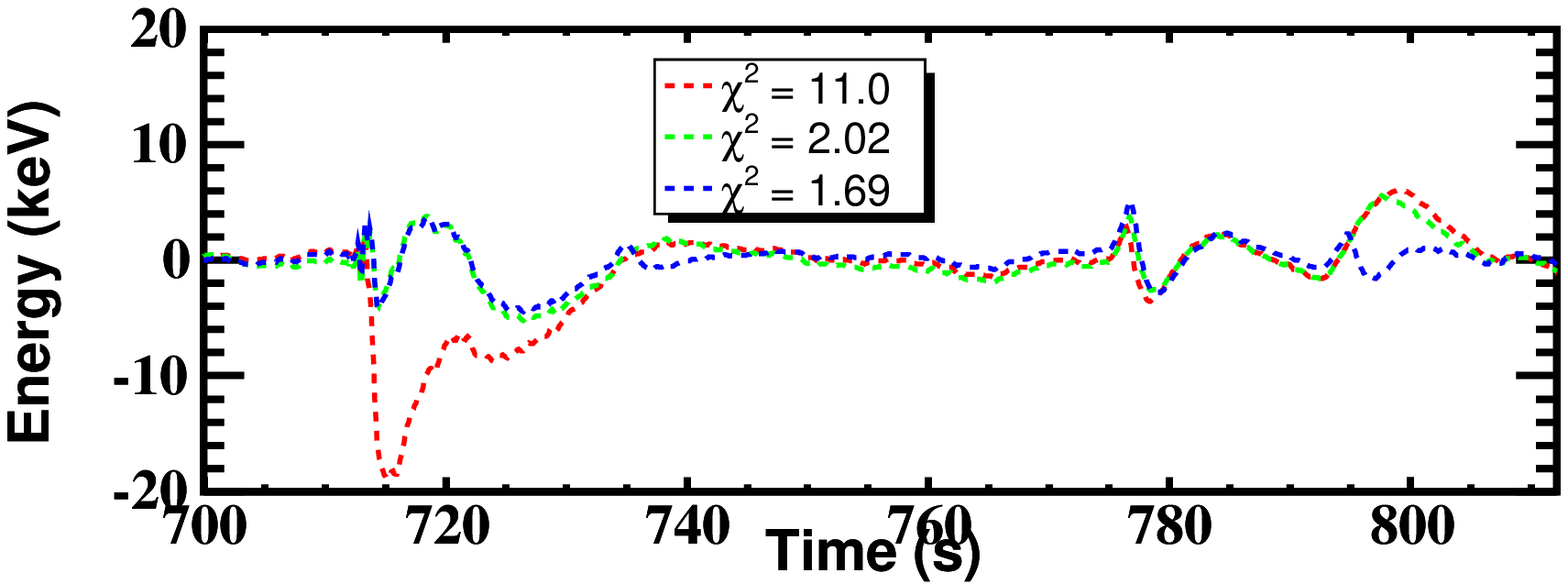}}
\caption{From top to bottom the first three plots present the data (in black), the best fit the
data (in red) and the residuals (in green) for
consecutive iterations of the algorithm with reduced
$\chi^2$ values of, $11$, $2.02$ and $1.64$ respectively.
  Vertical solid lines (in blue) show the position and the amplitude of the fitted peak for each
iteration. 
On the bottom plot, we represent the residuals for the
above three iterations in red, green and blue respectively. 
The amplitude of the signal is given in   keV, see \cite{electrons} for details.}
\label{fig:dataprocessing}
\end{center}
\end{figure}

These three steps are repeated until no peak at 5$\sigma$ is found in the residual data
meaning that we reached the noise level. In general we need about 10 iterations
per piece of data.  To avoid the algorithm to diverge we stop the iterative procedure
when more than 20 iterations are performed and that piece of data is not used
for further analysis (less than 1 \% of the data). 
At each iteration the position, the amplitude and the signal to noise ratio 
are recorded for each detected peak.
We characterize the quality of the fit using a $\chi^2$ test. 
The fit to the data with the smallest reduced $\chi^2$ is chosen as best fit.
Such a procedure allows us to have access to the peaks with the smallest
amplitudes, in which we are particularly interested.\\

As an illustration of the above method, we present in   
figure~\ref{fig:dataprocessing} the iterative analysis of
a sample of raw data corresponding to 110 seconds of
acquisition time. From top to bottom the first three
plots trace the data (in black), the best fit to the
data (in red) and the residuals (in green) for
consecutive iterations of the algorithm with reduced
$\chi^2$ values of $11$, $2.02$ and $1.64$ respectively.
On the bottom plot, we represent the residuals for the
above three iterations in red, green and blue respectively.
We can clearly observe from those figures how the fit improves
from one iteration to another. Further, by comparing the second and
third plots, we remark that the peak at 800 s is identified by the
algorithm only in the third iteration. This is because the residuals 
  at the end of iteration 2 have finally decreased to lower than the amplitude 
 of the 800 s
peak. Notice that only iteratively we can achieve such a precision on
the fit and therefore recover the lowest amplitude peaks in the data.

\section{Application to simulated data  \label{sec:appli.simu}}
 
We present in this section the application of the above analysis method
to the simulated MACHe3 data. We characterize
the quality of the reconstruction of the signal 
using as parameters both the efficiency of detection
and the contamination from spurious detections.
\subsection{Simulation of time ordered data  \label{sec:simu}}
To estimate the efficiency of the analysis method presented in the previous 
section, we have produced fake time ordered 
data from physical simulations using the Geant4 package \cite{g4}. For these simulations, 
we fully reproduced the instrumental set-up taking particular care of the detector geometry 
as well as of the materials  and we included 
the physical processes involved. More precisely, the G4EMLOW2.3 package has
been used to perform an adequate and accurate treatment of low energy 
electromagnetic interactions. 
As discussed before, the expected events in the 
data are related to cosmic muons, $\gamma$-rays 
and electrons from the $^{57}$Co source. For each contribution, the simulation
allows to compute its energy distribution. $\gamma$-rays at $\rm 14.4$, $\rm 122$ and $\rm 136 \ keV$ 
are emitted isotropically from the $\rm ^{57}Co$ source, assuming an activity of $\rm 0.06 \ Bq$.
Both conversion and Auger electrons are taken into account.
The energy corresponding to each line
is weighted to respect its intensity deduced from the decay   
scheme of the $\rm ^{57}Co$ source.

The simulated timeline for each cell is obtained as follows. 
1) The amplitude of each peak in the timeline is randomly chosen from 
the simulated energy distribution of the events. Indeed, the rates of each particle is recovered in the
simulated time line.
2) These amplitudes are then convolved by the estimated reference peak as shown on
Fig. \ref{fig:referencepeak}.
3) The peaks are uniformly distributed on the timeline so that
we can reproduce the pile-up observed in the data.
4) White noise is eventually added to the simulated timeline.
The noise as observed in the experimental timeline can be 
in a first approximation considered as white noise with a dispersion of $\rm 1-2 \ keV$.
However notice that in here we do not take into account more complex 
structures in the noise nor parasitics in the data  such as microvibrations.

\begin{figure}[!hb]
\begin{center}
\mbox{\hspace{-0.4cm}\includegraphics[scale=0.5]{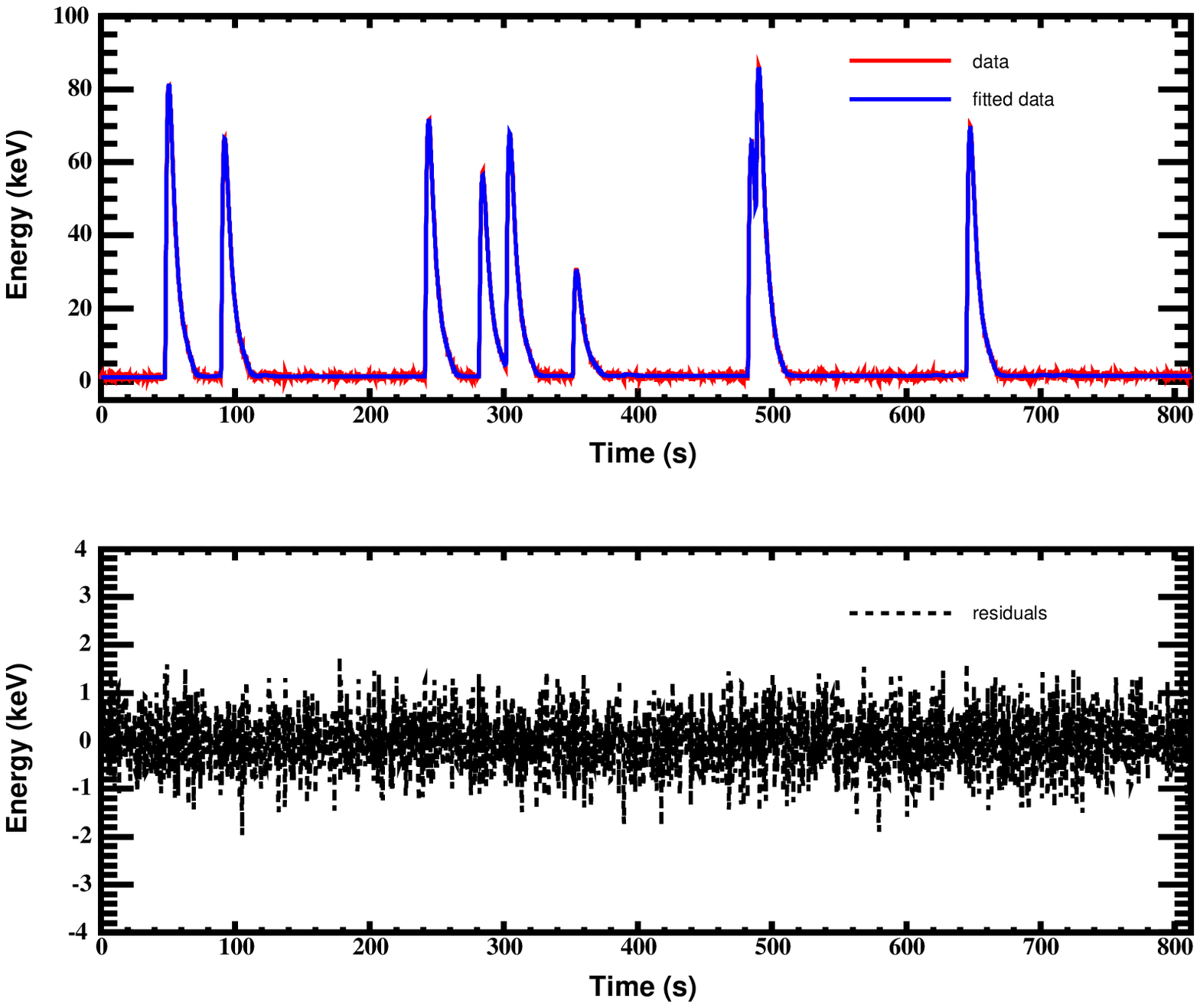}}
\mbox{\hspace{-0.4cm}\includegraphics[scale=0.5]{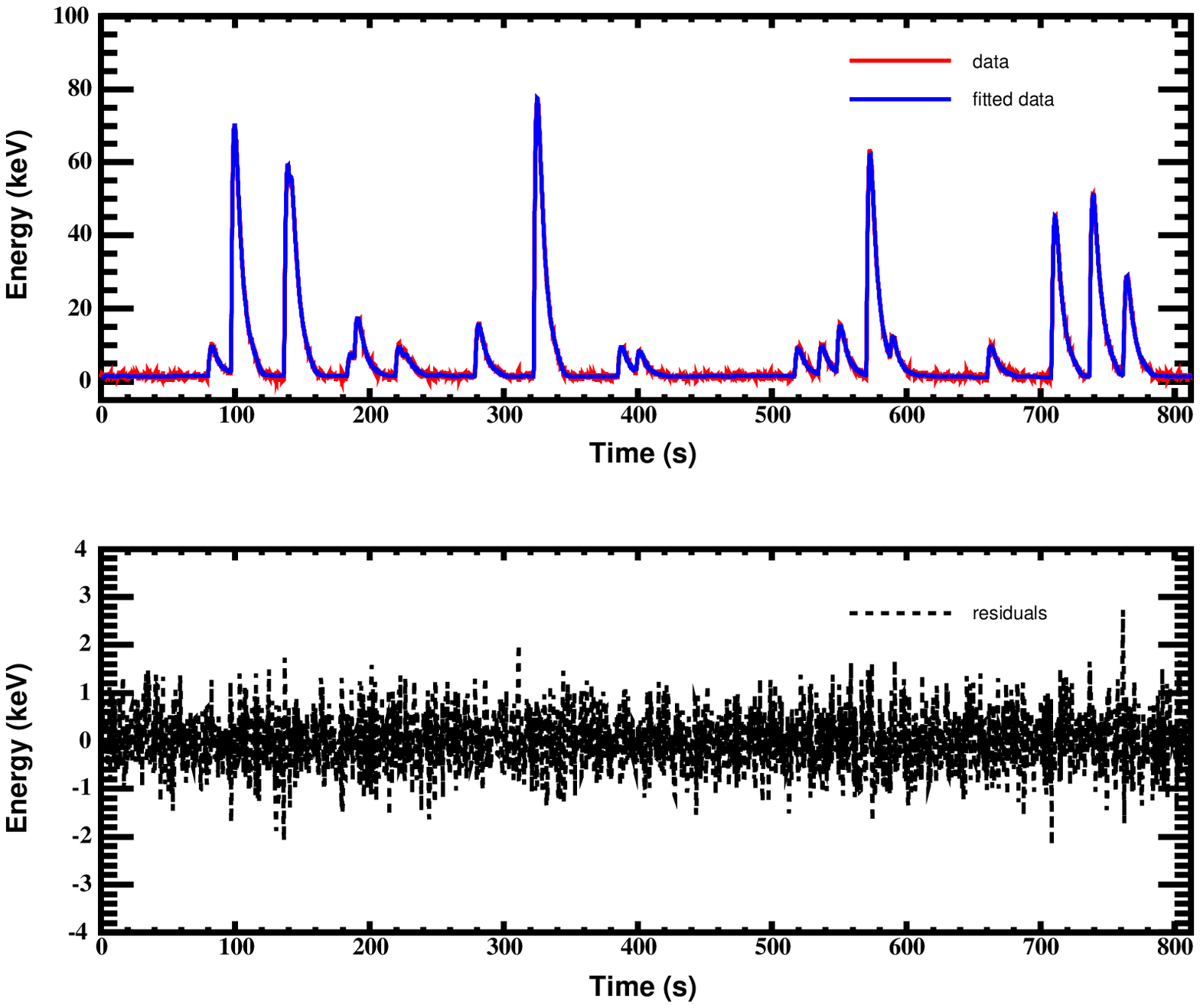}}
\caption{From top to bottom, 
simulated data sample (red solid line) for A and B set of simulations respectively.  
The best fit to the simulated data is presented (solid blue line) as well as the residuals (dotted line).
Residuals in each case are presented on an expanded scale below (black dashed line).}
\label{fig:todcellAB}
\end{center}
\end{figure}
\begin{figure*}[thb]
\begin{center}
\mbox{\includegraphics[scale=0.35]{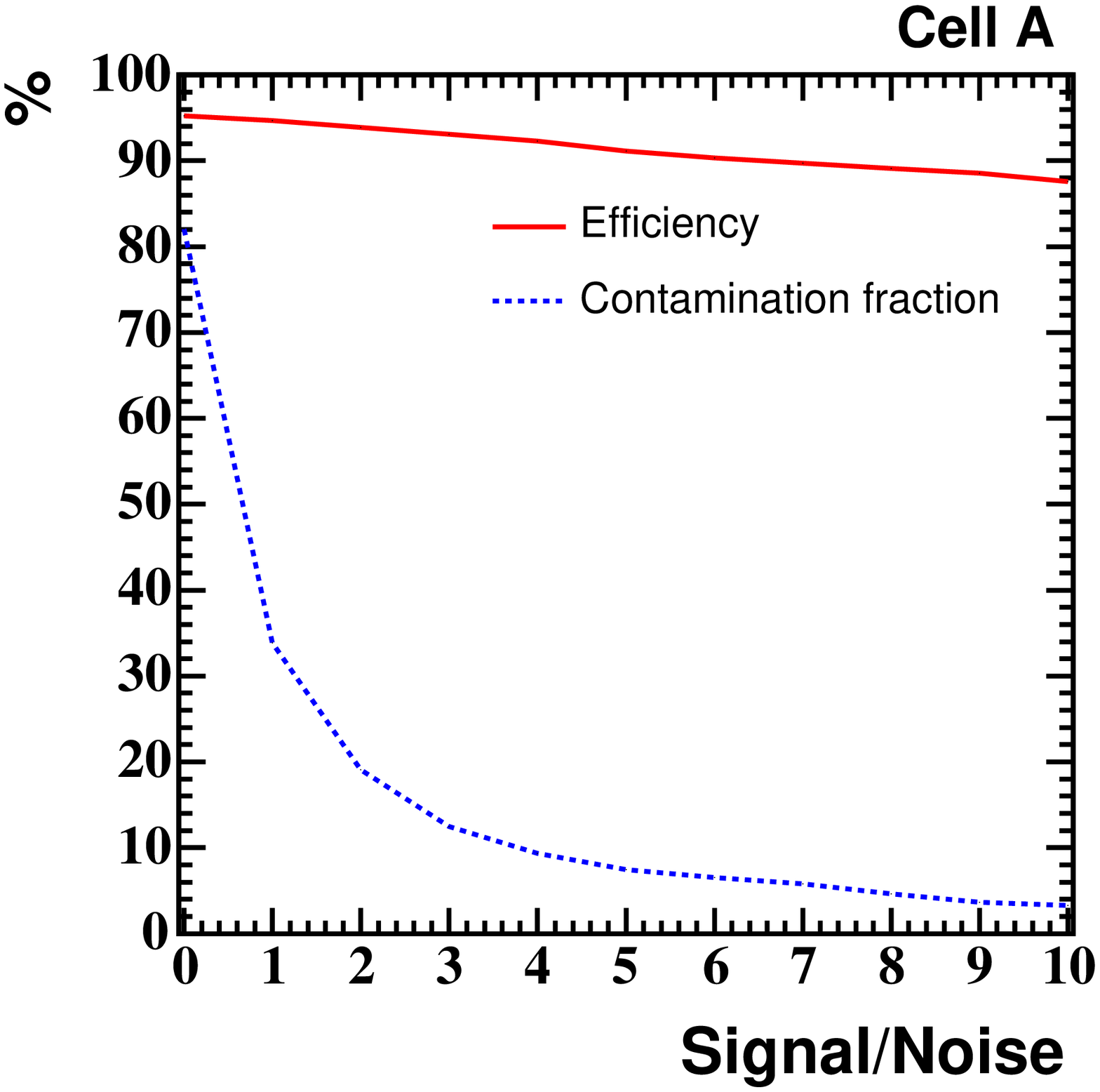}
\includegraphics[scale=0.35]{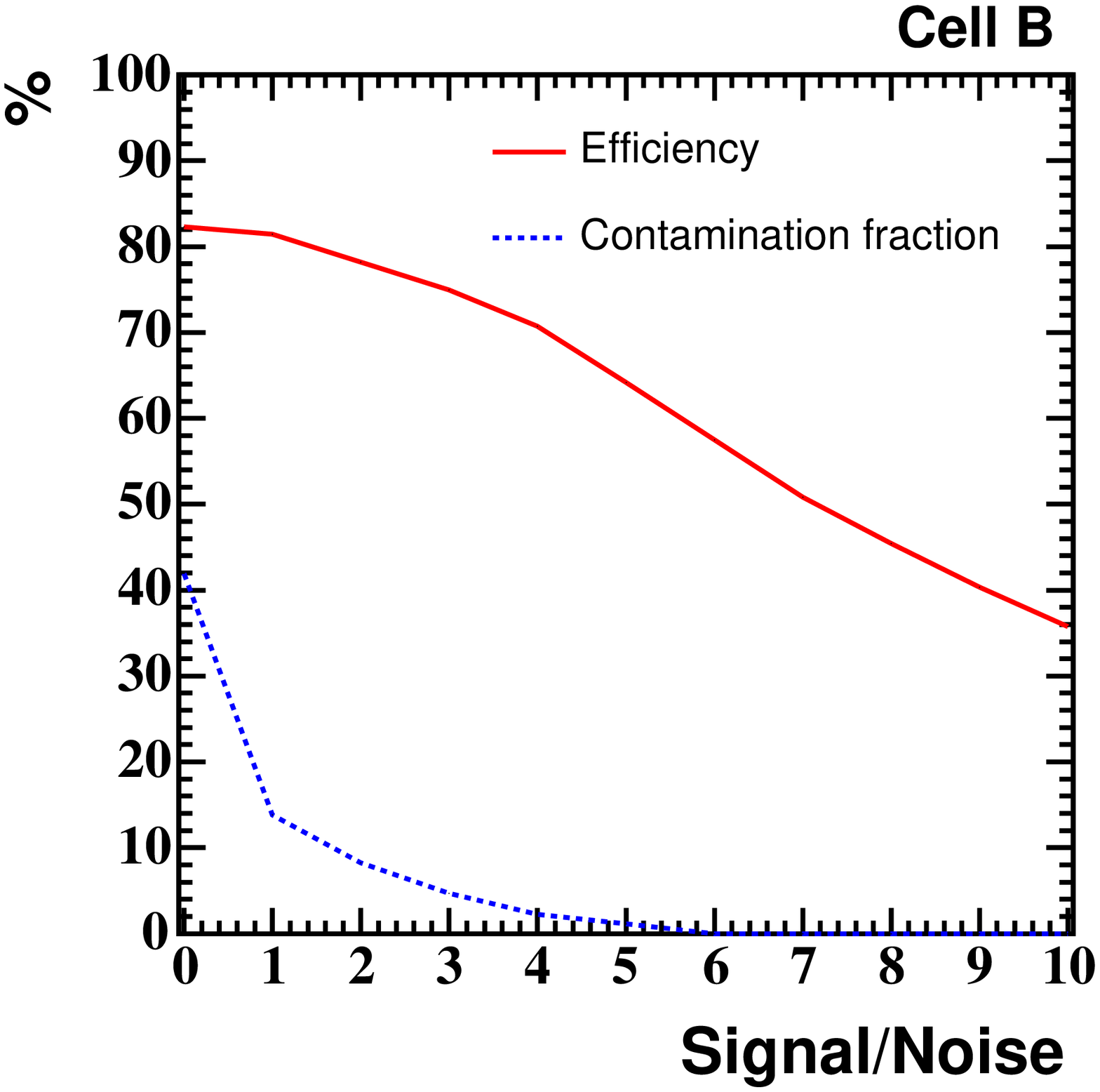}}
\caption{From left to right, global detection efficiency (in red solid line) and   contamination fraction 
(in blue dashed line) as a function of cut in the $\rm S/N $ ratio 
for the A and B set of simulations respectively (see text for details).}
\label{fig:globaleff}
\end{center}
\end{figure*}
\subsection{Reconstruction of the simulated data}
As a first step we concentrate in the analysis of a sample of simulated data of
about $\rm 819 \ s$. Figure~\ref{fig:todcellAB}, from top to bottom, shows in red
this data sample after denoising and baseline subtraction 
for two different sets of simulations A and B respectively. 
  The fit to the data is presented in blue and the lowest figure presents the residuals between data
and fitted data. The r.m.s. of the residuals is lower than 1 keV.

In the A simulations we include cosmic muons and $\gamma$-rays 
to reproduce the features of the MACHe3 A cell. The
B simulations include in addition the simulated conversion electrons
for \Co source to mimic the data in the MACHe3 B cell.
We plot in blue the best fit to the data and in dashed black line the residuals after subtraction of the
latter.  

In the case of the A simulations, the input timeline consists of 9 events. After analysis 
we recover 23 peaks among which 9 are considered as detected (meaning S/N $>1$) 
and 14 as non-detected (meaning $\rm S/N <1$). By imposing this detection condition,
we reject all spurious detections. Further, all the detected peaks present high $\rm S/N$ 
($\rm \geq 39$) as could  be expected since the simulated timeline does only 
contain large amplitude peaks corresponding to
the $\mu$ contributions with a relatively low pile-up rate. 

In the case of the  B simulations, the input timeline consists of 25 peaks with amplitudes 
ranging from 5 to 85 keV corresponding to the $\mu$, $\gamma$ and  $e^-$ contributions). 
Imposing the simple detection condition, $\rm S/N\geq 1$, 19 peaks are recovered 
from the best fit timeline and we have no spurious detection. We can therefore
consider that the rest of detected peaks represents the noise in the detection.
In this particular case, just considering only peaks with $\rm S/N\geq 1$ is
enough to remove all the noise contribution. However, in the general case
a careful study has to be performed to identify which $\rm S/N $ cut is
needed to have a negligible noise contribution. \\

From these results, we clearly observe that
the efficiency of reconstruction of the input signal will depend 
very much both on the amplitude of the peaks and on the
level of pile-up. Therefore, to fully characterize the quality
of the method we consider in the following, 
1) the efficiency of detection, defined as the number of reconstructed peaks 
divided by the total number of input peaks, and,
2) the   contamination fraction, defined as the number of spurious detections divided
by the total number of detected peaks.  Both quantities have been calculated as a function of S/N.


\subsection{Global analysis of the quality of the method}

We have performed a global analysis of the quality of reconstruction of the method based on 
$\rm 15.5 \ h$ of simulated data for the A and B cells.
Figure~\ref{fig:globaleff}
shows from left to right the efficiency of detection, in red, and the
  contamination fraction, in blue, as a function of the cut in the $\rm S/N $ ratio
\footnote{A cut in the $\rm S/N $ ratio of 5 means that we only consider in
the analysis reconstructed peaks with   $\rm S/N \ge 5$.}
of the detected peaks for the A and B cells respectively. 

For the A set of simulations,
we observe that the efficiency of detection is roughly constant with
the $\rm S/N $ cut. However the  contamination fraction quickly decreases
and, in particular, it is under 5 \% for  $\rm S/N$ cuts of 5 or above.

For the B set of simulations, the efficiency of detection decreases significantly for
large values of $\rm S/N $ cut with a turnover between 4 and 5. This is due to the fact that
the low amplitude peaks and specially those of a few keV are difficult
to fit with a background noise level of about 1 keV but also to
the high level of pile-up which reduces the discrimination of
the algorithm. With respect to the latter, it can be observed that 
for some of the pile-up peaks, the best fit to the data indicates more
peaks   than  are really  in the data. 

From the previous results we can conclude that a $\rm S/N$ cut of 5 would be a
good compromise in terms of low   contamination fraction and sufficiently high
detection efficiency for both the A and B set of simulations. Indeed, the  contamination fraction  would be less than 5 \% for the A simulations and negligible (at the accuracy
of our simulations) for the B simulations. The efficiency of detection would be
of the order of 90 \% for the A simulations and above 60 \% for the B simulations. 
  Further, it can be noted that the   contamination fraction  is higher in the cell A than in cell B 
due to the smaller number of total detected peaks.

\subsection{Evolution of the reconstructed simulated spectrum with the $\rm S/N$ cut}

\begin{figure}[t]
\begin{center}
\includegraphics[scale=0.65]{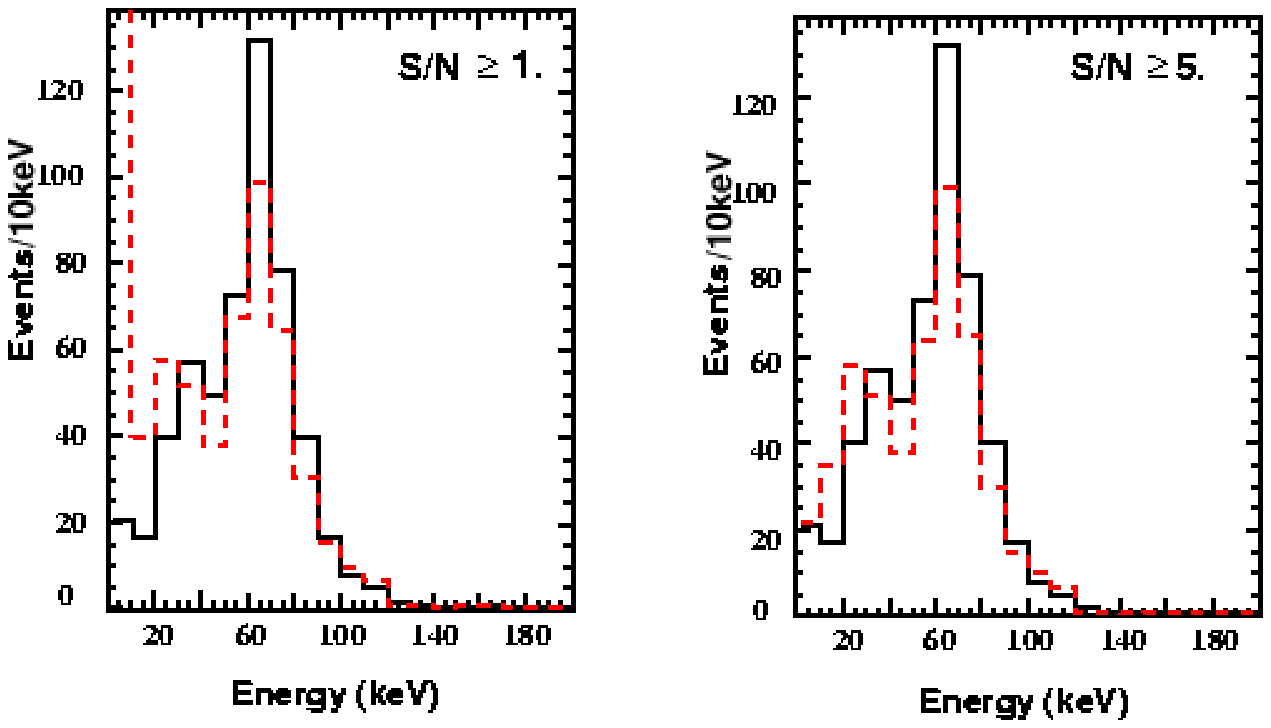}
\includegraphics[scale=0.65]{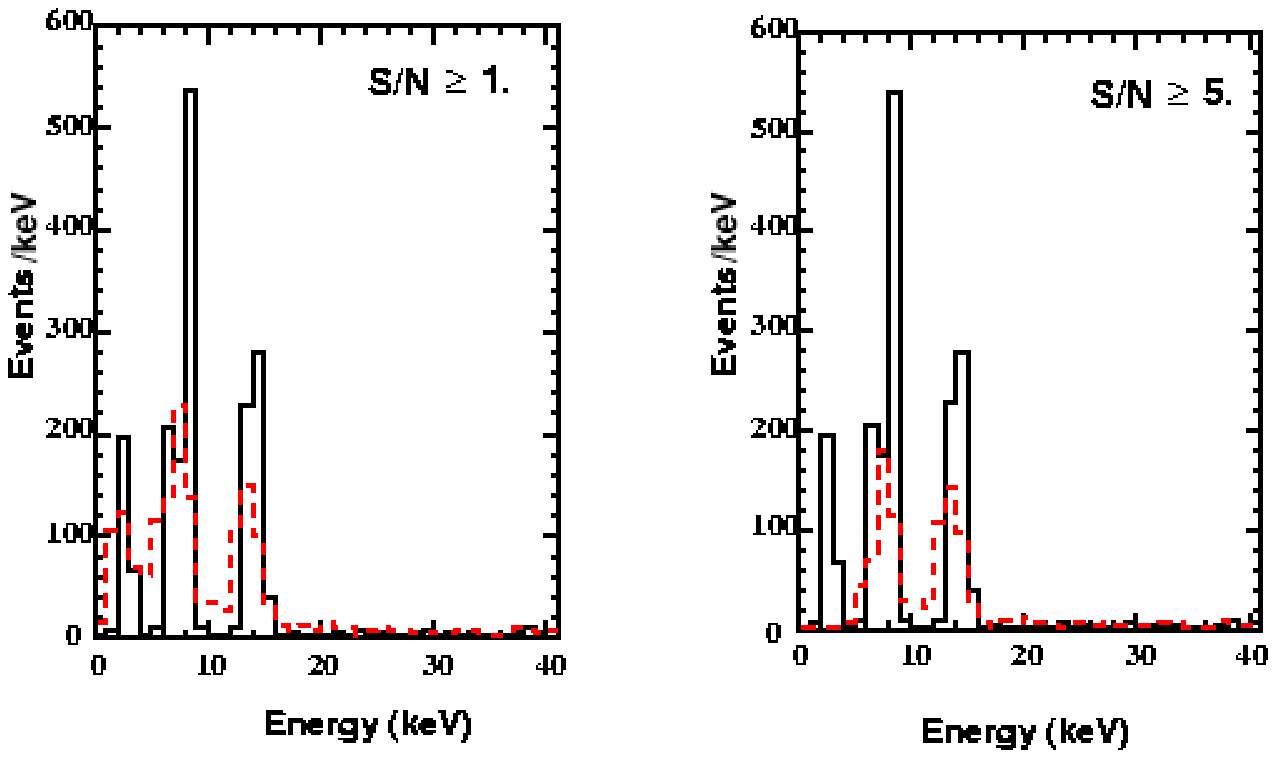}
\caption{In the top and  bottom panels, we plot, in red, the reconstructed simulated spectrum 
for the A set of simulations in the energy range $[0,200]$ keV  
and for the B set of simulations in the energy range $[0,40]$ keV  
for $\rm S/N$ cuts of 1 (left plot) and 5 (right plot). 
The input spectrum is overplotted in solid black line.}
\label{fig:evol}
\end{center}
\end{figure}

To complete the global picture presented above, we concentrate now in the
analysis of the quality of the reconstruction of the input energy spectrum of the signal.
For this purpose we show in figure~\ref{fig:evol},  the reconstructed spectrum
in red and the input spectrum in black for the A (top panel) and B (bottom panel) simulations
for values of the $\rm S/N$ cut of 1 (left plot) and 5 (right plot). 

In the case of the A set of simulations, for a $\rm S/N$ cut of 1, the reconstruction is very
good for energies above 20 keV but below we clearly observe spurious signals.
When assuming a $\rm S/N$ cut of 5 the contribution of the spurious signal 
is removed significantly at low energies and the quality of the reconstruction is maintained
above 20 keV. The differences observed between the reconstructed and
the input spectra comes, on one hand, from the fact that the algorithm does not have 100 \% 
efficiency and, on the other hand, by the errors on the determination of the amplitude
of peaks which tend to smear out the spectrum.

In the case of the B set of simulations, the analysis is more complex. For energies above 20 keV
we recover the same behavior than for the A simulations as the signal is dominated by
the cosmic $\mu$ contribution. At energies between 1 and 5 keV, and 
for a $\rm S/N$ cut of 1 we clearly observe
the contribution from spurious signals which is of the same order of that
observed for the A simulations. Although a significant contribution from the spurious
signal is still present we can clearly distinguish the three energy lines
corresponding to electrons of $\rm \sim 2 \ keV$ (pile-up of L-shell Auger electrons from \Co ), $\rm 7.3 \ keV$ 
and $\rm 13.6 \ keV$ (conversion electrons). For a $\rm S/N$ cut
of 5 the spurious signal is roughly completely removed but the electron line at $\rm \sim 2 \ keV$ is lost. 
However, the lines at 7.3 and 13.6 keV
are preserved although smeared out in the same way as discussed above
for the A simulations.

\subsection{Efficiency of the reconstruction for low energy events. \label{sec:efficacity1-5}}
\begin{figure}[t]
\begin{center}
\includegraphics[scale=0.4]{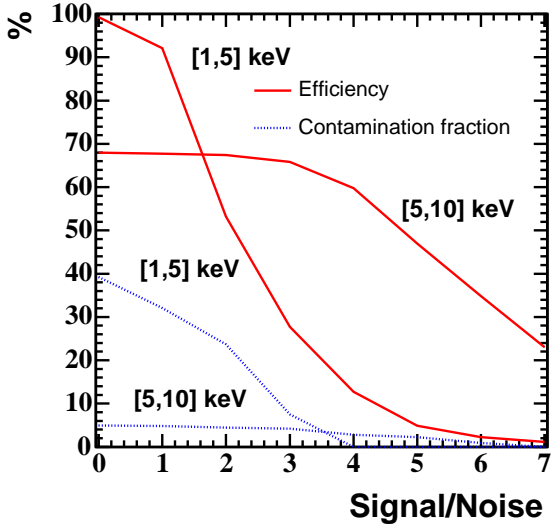}
\caption{Detection efficiency   (red solid line)  and   contamination fraction  
 (blue dashed line),  
in the energy ranges 1-5 keV and 5-10 keV as function of 
$\rm S/N$ cut for the B set of simulations.}
\label{fig:effAB}
\end{center}
\end{figure}

In the analysis of the B cell data presented in the following section
we are particular interested in the reconstruction
of the very low amplitude events in the energy range from 1 to 16 keV (range of 
energies for conversion electrons).
As discussed above for our choice of $\rm S/N$ cut of 5 and for the B simulations we reduce significantly
the contamination from spurious signals but also the efficiency of reconstruction
of the low energy events. This can be clearly seen in figure~\ref{fig:effAB}
where we represent the detection efficiency, in red, and the   contamination fraction, in blue, in the energy ranges 1-5 keV and 5-10 keV as function of 
$\rm S/N$ cut for the B simulations. In the energy range from 5-10 keV we observe
that the detection efficiency is roughly constant up to a $\rm S/N$ cut of five
and above 60 \%. By contrast in the range from 1-5 keV the detection
efficiency decreases dramatically for $\rm S/N$ cuts above 2 being of the
order of 5 \% at a $\rm S/N$ cut of 5. Finally, the   contamination fraction
is well below 5 \% for both energy ranges. We conclude that
for a $\rm S/N$ cut of 5 we can not have access to events at energies below
5 keV.

\section{Application to MACHe3 data \label{sec:appli.mache3}}
We present here the application of the method described in section\ref{sec:anal} to the MACHe3 data.
We focus on the reconstruction of the low energy conversion electron spectrum from the B cell data.
In addition, we describe the background contamination from cosmic muons as estimated from the A cell data (without
the \Co source).

\subsection{Results on the raw data}
\begin{figure}[!ht]
\begin{center}
\mbox{\hspace{-0.5cm}\includegraphics[scale=0.51]{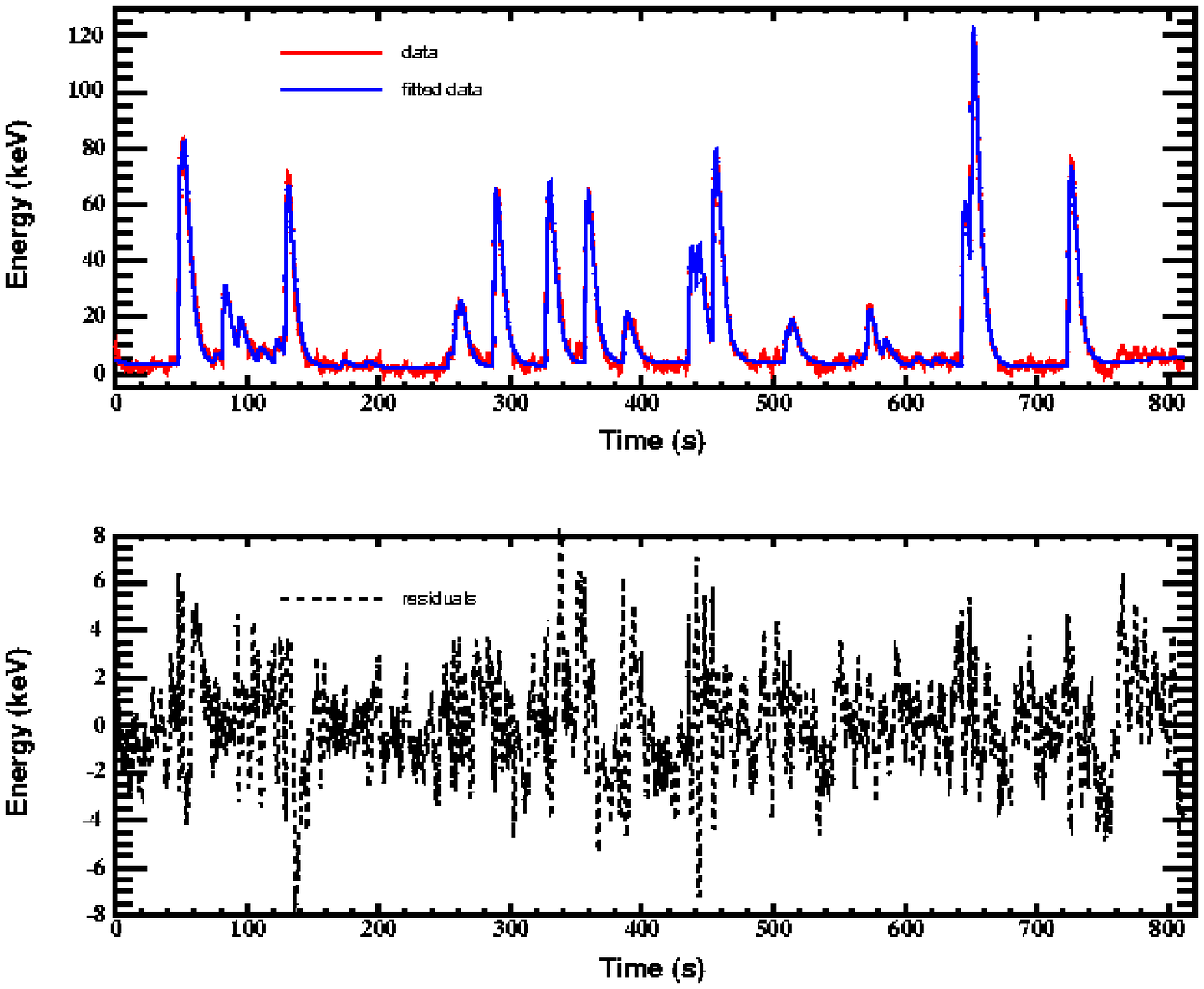}}
\mbox{\hspace{-0.5cm}\includegraphics[scale=0.51]{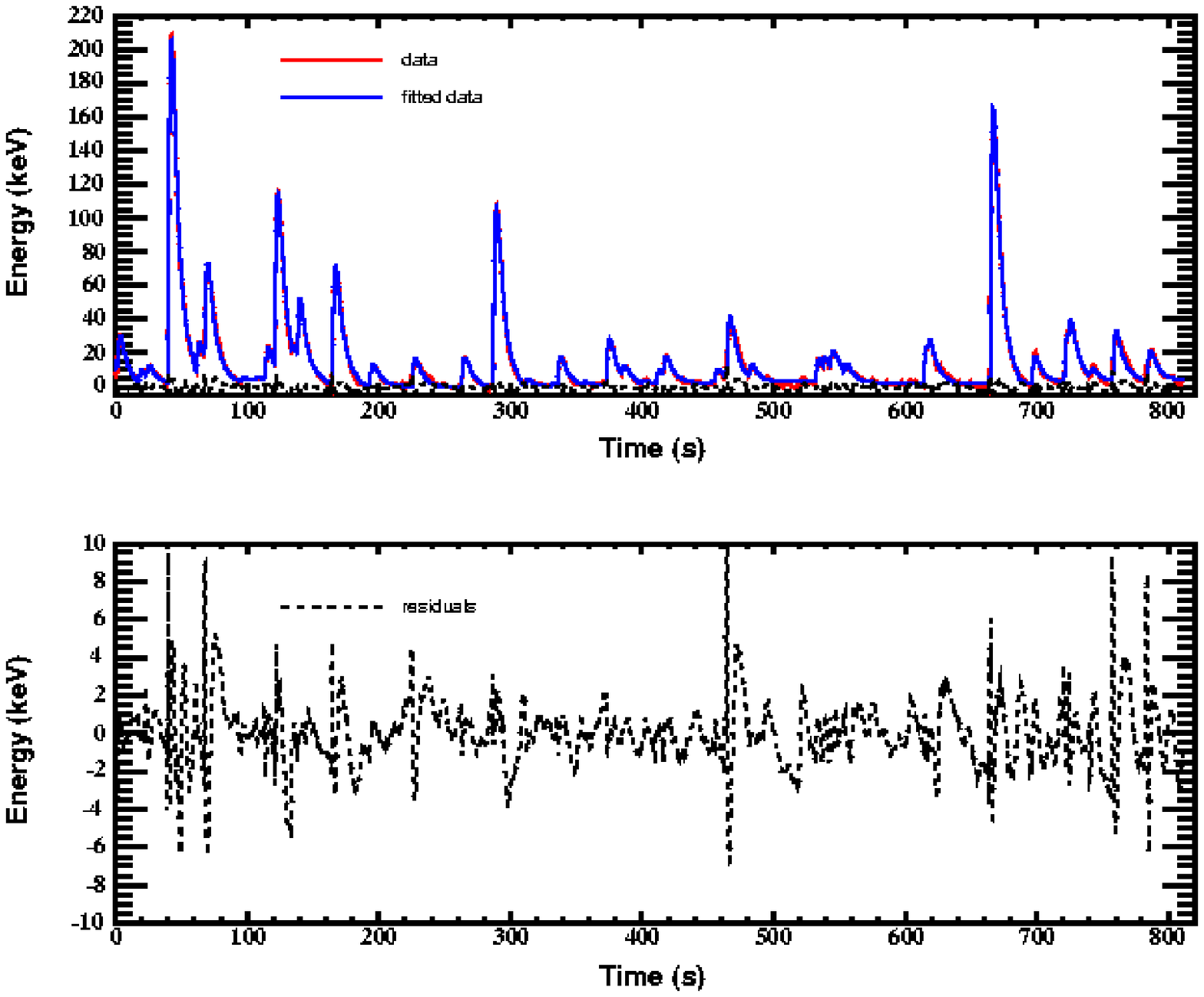}}
\caption{Sample of raw data (red) as well as best fit to the data 
(blue) for the A  (upper
panel) and B (lower panel) cells.   In each case, residuals are shown below (dotted line).}
\label{fig:rawdataAB}
\end{center}
\end{figure}

In figure \ref{fig:rawdataAB},
we present the main results of the analysis method on a sample of
experimental data for cells A (upper plots) and B (bottom plots).
We observe that the level of pile-up is much more important
in the B cell 
due to the presence of the low energy Auger and conversion electrons emitted
by the source.
For both cells, the peaks corresponding to cosmic muon interactions are clearly visible
in the data, as they present larger amplitudes, up to $\sim$ 100 keV. For the B cell, the detected low amplitude
 peaks correspond to the low energy electrons. 
We plot, in blue, the best fit to the data and residuals in black.
We observe that the fit is as good as the one for the simulations presented in section~\ref{sec:appli.simu}.
It is important to notice   the noise
in the raw data is more complicated than in the simulated noise.
However, the algorithm with about ten iterations, permits to access to the lowest amplitudes and
the final residuals are at the level of the noise measured in data sample without peaks.
Despite the high pile-up rate, the method permits to
recover the low energy events we are interested in.

\subsection{Low energy conversion electron spectrum}
\begin{figure}[h]
\begin{center}
\includegraphics[height=6cm,width=8.5cm]{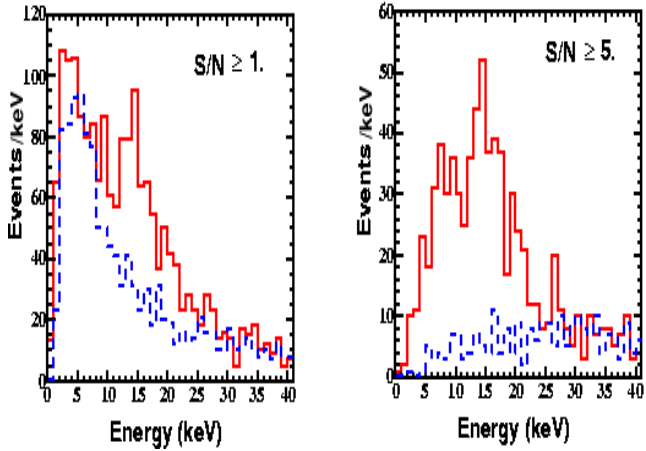}
\caption{From left to right, we overplot the spectra for the A (blue   dashed line) 
and  B (red  solid line) cells in the $[0,40]$ keV energy range for values
of the $S/N$ cut of 1 and 5 respectively.}
\label{fig:evoldata}
\end{center}
\end{figure}
The experimental energy spectrum has been analyzed in detail in the $[0,40]$ keV
energy range for both cells.
The evolution of the spectrum with respect to the signal to noise cut is presented on figure~\ref{fig:evoldata}.
We plot in blue and red the energy spectrum of the A and B cells respectively. The left (resp. right) plot corresponds to a cut in the 
$S/N$ of 1 (resp. 5).\\
From the left plot, we conclude that, for a $S/N$ cut of 1, the reconstruction of the spectrum of the 
low energy conversion electrons is biased by spurious signals in the range from 0 to $\rm \sim 20 \ keV$. 
This is clearly visible from the spectrum of the A cell which increases with decreasing energy.  
The same features were observed in the analysis of the simulated data in section~\ref{sec:appli.simu}. 
For a signal to noise cut of 5, the spurious signals are   mostly removed from the
spectrum,   at a level better than 95\%.
 This is   suggested by the spectrum of the A cell which decreases toward the lowest energies
as expected both from the decrease of efficiency of detection and from the negligible
contribution from spurious signals discussed in section~\ref{sec:efficacity1-5}. 
Notice in here  that the activity of the source was not accurately previously known and was estimated from this analysis. Hence, only
qualitative comparison between simulation and data can be done at this stage.
By comparing with
figure~\ref{fig:evol}, we can highlight that the lines for the conversion electrons at 7.3 and 13.6 keV are
recovered. The reconstructed relative intensity between these two lines does
not match what observed in the simulations. This is due to the fact that for the data spectrum
the line at $\sim 13$ keV is the combination of the conversion electron line at 13.6 keV and
of an extra line from the Auger electrons from the Au foil on which the source was deposited.
Equally, two extra lines are present, corresponding to the physical pile-up of the conversion and 
Au-Auger electron lines.  The contribution from the Au-Auger electrons was not taken into account in the 
simulation spectrum.\\
A more detailed analysis and physical interpretation of this spectrum as  
the detection of low energy conversion electrons in \hetrois at ultra low temperature can be found in \cite{electrons}.

\section{Conclusion\label{sec:conc}}
We present in this paper the design and implementation of a data analysis
method to reconstruct highly piled-up bolometric peak-like signals from
time ordered data.  This method consists of three main steps: 1) wavelet
based denoising to remove the high frequency noise in the data, 2) reconstruction
and subtraction of the low frequency drift in the data from the local minima to
avoid the bias in the amplitude of the reconstructed signal, 3) an iterative detection and
fitting procedure. The detection is performed in the derivative of the
residuals and for the fit we use a reference peak directly extracted from the data.
For each reconstructed signal we compute its time position, amplitude and
$S/N$ ratio of detection defined as the amplitude of the peak over the maximum
of the residuals in the interval of detection.

The method has been applied to simulated data of the MACHe3 prototype experiment
in two configurations: 1) large amplitude signals (from 10 to 120 keV) with low level of pile-up (cell A)
and 2) small (few keV) and large amplitude signal with high level of pile-up. 
From this we have proved that the method is able to recover the lowest-amplitude   simulated 
signals of few keVs for a rms noise of 1 keV. Considering a $S/N$ cut of 5,
the efficiency of detection in the A cell (resp. B cell) configuration is above 90 \% (resp. 60 \%) and
the contamination by spurious signals below 5 \%. (resp. 1 \%).

For the B cell and for a $S/N$ cut of 5 the efficiency of detection of the low 
amplitude  simulated  signals between 5-10 keV is above 50 \% with
negligible spurious contribution. For the same $S/N$ cut in the range from 1 to 5 keV, although
we have no spurious signal, the efficiency of detection is too low because of the 
high level of pile-up. This does not mean that we are not able to detect low
amplitude  simulated  signals but that we are limited by the pile-up in this analysis. 
For example, if we consider a $S/N$ cut of 2 the efficiency of
detection is above 60 \% but we have to tolerate about 20~\% of spurious signal.
In the case of dark matter detection, in underground laboratories, there would be significantly less pile-up and therefore
the low amplitude signals from 1 to 5 keV could be detected and reconstructed.

We have also applied the method to the raw data
of the MACHe3 prototype experiment, in order to demonstrate for the first time
the possibility of measure low energy events  (5-10 keV) in \hetrois 
at ultra-low temperature ($ \rm 100\ \mu K$)~\cite{electrons}.
Indeed, we were able to detect events corresponding to conversion electrons at 7.3 and
13 keV coming from \Co source embedded in the prototype and to reconstruct their energy
spectrum.

The method, although developed for the analysis of the MACHe3 data, is only based
on the properties of the time ordered data. As ongoing and forthcoming
Dark Matter experiments present the same features in their data, this procedure 
could be of general interest.

\begin{acknowledgements}
 We acknowledge D. Yvon for comments and fruitful discussions.

\end{acknowledgements}

\end{document}